\documentclass[twocolumn,showpacs,amsmath,floatfix,prb]{revtex4}
\usepackage{graphicx,times}
\begin{document}

\title{ Second-order Optical Response from First Principles}

\author{S. Sharma}
\email{sangeeta.sharma@uni-graz.at}
\author{C. Ambrosch-Draxl}
\affiliation{Institute for Theoretical Physics, Karl--Franzens--Universit\"at Graz,
Universit\"atsplatz 5, A--8010 Graz, Austria.}

\date{\today}

\begin{abstract}

We present a full formalism for the calculation of the linear and second-order optical response for semiconductors and insulators.
The expressions for the optical susceptibilities are derived within perturbation theory. As a starting point a brief background of the single and
many particle Hamiltonians and operators is provided. As an example we report calculations of the linear and nonlinear optical properties of 
the mono-layer InP/GaP (110) superlattice.
The features in the linear optical spectra are identified to be coming from various band combinations. The main features
in the second-order optical spectra are analyzed in terms of resonances of peaks in linear optical spectra. With the help of the
strain corrected effective-medium-model the interface selectivity of the second-order optical properties is highlighted.     

\end{abstract}

\maketitle

\section{INTRODUCTION}

A material interacting with the intense light of a laser beam responds in a "nonlinear fashion". Consequences of this are a number of peculiar 
phenomena, including the generation of optical frequencies that are initially absent. This effect allows the production of laser light at wave
lengths normally unattainable by conventional laser techniques. So the application of nonlinear optics (NLO) range from basic research to
spectroscopy, telecommunications and astronomy. Second harmonic generation (SHG), in particular, corresponds to the appearance
of a frequency component in the laser beam that is exactly twice the input one.
SHG has great potential as a characterization tool for materials, because of its sensitivity to symmetry. Today SHG is widely applied for studying 
the surfaces and interfaces because it requires an inversion asymmetric material. For materials with bulk inversion symmetry SHG is only allowed 
at surfaces and interfaces. This makes SHG a powerful surface selective technique. SHG in conjunction with Kerr and Faraday rotation\cite{qiu00} 
is used as an excellent  tool for studying magnetic surfaces. In case of embedded interfaces this technique gains extra weight when an intense 
laser is used which is capable of penetrating deep into the material and no direct contact with the sample is needed. 

In the case of linear optical transitions an electron absorbs a photon from the incoming light and makes a transition to the next higher unoccupied 
allowed state. When this electron relaxes it emits a photon of frequency less than or equal to the frequency of the incident light (Fig. 1(a)). 
SHG on the other hand is  a two photon process where this excited electron absorbs another photon of same frequency and 
makes a transition to yet another allowed state at higher energy. This electron when falling back to its original state emits a photon
of a frequency which is two times that of the incident light (Fig. 1(b)). This results in the frequency doubling in the output. 

\begin{figure}[ht]
\centerline{
\includegraphics[scale=0.5,width=\columnwidth,angle=0]{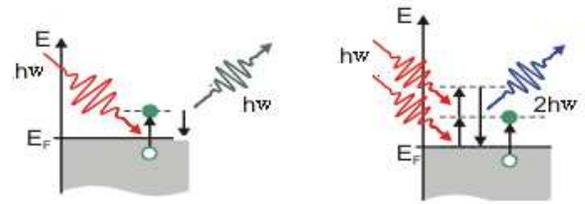}
}
\caption{ Schematic representation of a (a) linear optical transition and (b) second harmonic generation. }
\end{figure}
  
In order to extend the use of NLO for understanding the properties 
of surfaces and interfaces and for extracting maximal information from such measurements for non centero-symmetric materials, a more 
quantitative theoretical analysis is required. The calculation of the SHG susceptibility from first principles is an important 
but difficult task. The major work in this direction for semiconductors can be found in Refs. \onlinecite{sipe93,sipe96,segey98,sipe00} and 
for metals in Refs. \onlinecite{luce98,puStogowa93,hubner89,torsten02} and references there in. In this work, we present the formalism for calculating
the second order susceptibility $\chi^{(2)}(2\omega,\omega,\omega)$ for non magnetic semiconductors and insulators, within the independent particle 
approximation. The expressions formulated are amenable for numerical calculations using any band structure method and are computationally more 
efficient\cite{SL} than the previously presented expressions. \cite{sipe96,sipe93,sipe00,segey98} 

As an example, we present the linear and nonlinear optical properties of a InP/GaP superlattice (SL). Semiconducting strained SLs are potential materials
for applications in optical communications involving switching, amplification and signal processing. In particular III-V semiconductor hetero-structures 
and SLs have attracted a great deal of interest mainly due to the possibility of tailoring band gaps and band structures by variation
of simple parameters like superlattice period, growth direction and substrate material. Much of the  theoretical work done to understand the  
physical properties of SLs has been largely concerned with the understanding of the electronic band structure. For example, the effect of strain on 
the band gap, the band offset problem and the possibilities of engineering it as well as the interface energy and band structure
have been studied. \cite{franceschetti99,walle86,chris87,munoz90,agrawal99,dandrea91,tanida94,park93,kurimoto89,arriga91,franceschetti94,kobayashi96} 
The major theoretical work in the direction of NLO properties was done by Ghahramani {\it et al}. \cite{ghahramani92,ghahramani91-si,ghahramani90,ghahramani91}
They employed the non-self consistent linear-combination-of-Gaussian-orbitals (LCGO) method to calculate the band structures and optical 
properties of SLs. As an example, we present a fully self-consistent calculation of the nonlinear optical properties of the mono-layer 
InP/GaP (110) superlattice (SL). The structure in the linear optical spectra is identified from the
band structure of the material. The features in the second-order optical response are further analyzed in terms of resonances  of the
peaks in the linear optical spectra. With use of the strain corrected effective-medium-model (SCEMM), \cite{SL} and we identify the features in the 
optical spectra of InP/GaP coming from the SL formation. 
 
The paper is arranged in the following manner. In Section II we present the detailed formalism for calculating the SHG susceptibility. 
Section III deals with the linear and second-order optical response of an InP/GaP SL. 

\section{FORMALISM}

The aim of the present work is to obtain the expressions for the optical response of a material. The effect of the electric field vector 
${\bf E}(\omega)$ of the incoming light is to polarize the material. This polarization can be calculated using the following relation: 
\begin{eqnarray}\label{PEXP}
{\bf P}^a(\omega) = \chi^{(1)}_{ab}.{\bf E}^b(\omega)+\chi^{(2)}_{abc}.{\bf E}^b(\omega).{\bf E}^c(\omega)+\ldots
\end{eqnarray} 
In this expression $\chi^{(1)}$ is the linear optical susceptibility and $ \chi^{(2)}$ is the second order optical susceptibility. The higher order 
terms can also be calculated but in the present work we are only interested up to the second-order optical response. 
The formulae for calculating the SHG susceptibility, $\chi^{(2)}(2\omega,\omega,\omega)$, have been presented before.\cite{sipe96,sipe93,sipe00,segey98} 
In the present work we provide the detailed formalism. Before deriving the expressions for calculating the SHG susceptibility using perturbation 
theory, a brief background of the single and many particle Hamiltonians and operators is needed. This is presented in the following three sections. 

\subsection{THE HAMILTONIANS}

The optical response is treated within the independent particle approximation and the Hamiltonian in a.u. (i.e. $\hbar$
= $m$ = $e$ = 1) is written as  
\begin{eqnarray} \label{MPH}
H(t) & = & \sum_i {{({\bf p_i- K}(t))^2} \over 2} +V({\bf x}_i)
\end{eqnarray}
The subscript $i$ labels the electrons in the crystal at position ${\bf x}_i$. ${\bf p}$ is the momentum operator given by ${\bf p}_i = -i\nabla_i$. 
$V({\bf x})$ is the effective periodic crystal potential. ${\bf  K}(t)= {\bf A}(t)/c $ where ${\bf A}(t)$ is the vector potential of the external applied 
field. The macroscopic electric field is given by ${\bf E}(t) = -\dot{{\bf A}}(t)/c$. In the long wavelength limit the variation of this field over the 
distance of the lattice spacing is neglected. Eq. (\ref{MPH}) can be separated into a time independent (which may be implicitly time dependent) and explicitly 
time dependent parts as  
\begin{eqnarray}
H = H_0 + H_1 + H_2
\end{eqnarray}
with 
 \begin{eqnarray}
H_0 & = & \sum_i H_{0i} = {1 \over 2}\sum_i {\bf p}_i^2 +V({\bf x}_i)
\end{eqnarray}
\begin{eqnarray}
H_1(t)  & = &  -{\bf K}(t)\sum_i {\bf p}_i
\end{eqnarray}
\begin{eqnarray}
H_2(t)  & = & {1 \over 2} N{\bf  K}^2(t)
\end{eqnarray}
$N$ is the total number of electrons in the volume $\Omega$ of the crystal. In the long wavelength limit $H_2$ only introduces a time dependent 
phase factor for the wave functions and hence can be neglected. $H_1$ can be treated as a perturbation. The eigenstates of $H_0$ are given by
\begin{eqnarray}\label{TISE}
H_0\psi_n({\bf k,x}) & = & \omega_n({\bf k}) \psi_n({\bf k,x}) \nonumber \\
\psi_n {\bf (k,x}) &  =  & \Omega^{-1/2}u_n({\bf k,x})e^{i{\bf k.x}}
\end{eqnarray}
Next we consider the time dependent single particle Hamiltonian 
$H(t) = {1 \over 2} {({\bf p} -{\bf  K}(t))^2} +V({\bf x})$. The instantaneous eigenstates of this Hamiltonian are
\begin{eqnarray}\label{TDES}
\bar{\psi}_n({\bf k,x}) & = & \Omega^{-1/2}u_n({\bf k+K}(t),{\bf x})e^{i{\bf k.x}}
\end{eqnarray}
satisfying
\begin{eqnarray}\label{TDSE}
H(t)\bar{\psi}_n({\bf k,x}) & = & \omega_n({\bf k+K}(t))\bar{\psi}_n({\bf k,x}).
\end{eqnarray}
$\bar{\psi}_n({\bf k,x})$ is implicitly time dependent via ${\bf K}(t)$. If an orthonormal set $\bar{\psi}_n({\bf k,x})$ satisfies Eq. (\ref{TDSE})
at time $t$, then the same equation is satisfied at time $t+dt$ if
\begin{equation}\label{215}
i\hbar {d \over dt} \bar{\psi}_n({\bf k,x}) =  \sum'_m \bar {\psi}_m({\bf k,x}) \mu_{mn}({\bf k},t){\bf E}(t)
\end{equation}
for each n. If $\omega_m({\bf k+K})=\omega_n({\bf k+K})$ then
\begin{eqnarray}
{\bf E}(t).{\bf V}_{mn}({\bf k},t)=0. 
\end{eqnarray}
As shown Appendix A the above conditions require:
\begin{eqnarray}\label{mu}
{\bf \mu}_{mn}({\bf k+K},t) & = & {{{\bf V}_{mn}({\bf k},t)} \over {i\omega_{mn}({\bf k+K})}}
\end{eqnarray}
where
\begin{eqnarray}
V_{mn}({\bf k},t) & = &\int \bar{\psi}^*_m({\bf k,x})e^{-i{\bf K.x}}[-i\nabla](\bar{\psi}_n({\bf k,x})) \nonumber \\
&\times&e^{i{\bf K.x}}d{\bf x}
\end{eqnarray}

\subsection{THE OPERATORS IN SECOND QUANTIZATION}

We now introduce Fermionic raising and lowering operators $a^+_n$ and $a_n$ satisfying the anti-commutation relations $\{ a^+_n, a^+_m\} =\{ a_n, a_m\} = 0$ 
and $\{ a^+_n, a_m\}=\delta_{nm}$. Similar commutation relations are satisfied by  $b^+_n$ and $b_n$ which are Fermionic raising and lowering operators,
implicitly dependent on time via the wave function. The Hamiltonians in the second quantized representation are 
\begin{eqnarray}
H_0 & = &  \sum_{n{\bf k}} \omega_n({\bf k}) a^+_{n{\bf k}}a_{n{\bf k}} \\
H(t) &  = &  \sum_{n{\bf k}} \omega_n({\bf k+K}) b^+_{n{\bf k}}b_{n{\bf k}} \label{HOF}
\end{eqnarray}
A unitary transformation operator can be used for going from the implicitly time dependent operators $b_n$ and $b^+_n$ to time independent operators $a_n$ 
and $a^+_n$ as
\begin{eqnarray}
a_n & = & Ub_nU^+ \nonumber \\
a^+_n & = & Ub^+_nU^+
\end{eqnarray}
Here 
\begin{eqnarray} \label{uniop}
U & = & \sum_S |S> <\bar{S}| \nonumber \\
U^+ & = & \sum_S |\bar{S}> <S|
\end{eqnarray}
such that $|S>$ are states given by Eq. (\ref{TISE}), $|\bar{S}>$ are states given by Eq. (\ref{TDES}) and the sum runs over all the states. The 
importance of such a transformation will become clear in the next section (see for example Eq \ref{UTH} and the discussion after it). Since
our final aim is to calculate the linear and nonlinear optical response of the material another useful operator 
is the current density operator ${\bf J}$ which is given by
\begin{eqnarray}\label{Curden}
{\bf J} &  =  & {1 \over \Omega} \sum_i{\bf  p_i - K}(t)  \\ \nonumber
& = & {1 \over \Omega} \sum_{nm{\bf k}} 
b^+_{n{\bf k}}b_{m{\bf k}}{\bf V}_{nm}({\bf k+K}) \nonumber \\
{\bf J}' & = & \sum_{nm{\bf k}} a^+_{n{\bf k}}a_{m{\bf k}}{\bf V}_{nm}({\bf k+K}).
\end{eqnarray}
As shown in the Appendix B the current density is related to the response of the materials via polarization as
\begin{equation}\label{tot_curen_den}
{\bf J} = {\bf J_A} + {d {\bf P} \over dt}
\end{equation}
here ${\bf J_A}$ is the intraband current density given by Eq. \ref{ja} and ${\bf P}$, the effective polarization of the material given by
Eq. (\ref{PEXP}). In second quantization ${\bf P}$ can be written as
\begin{eqnarray}\label{Pol}
{\bf P} ={1 \over \Omega}\sum_{nm{\bf k}} b^+_{n{\bf k}}b_{m{\bf k}}{\bf \mu}_{nm}({\bf k},t) \nonumber \\
{\bf P}' ={1 \over \Omega}\sum_{nm{\bf k}} a^+_{n{\bf k}}a_{m{\bf k}}{\bf \mu}_{nm}({\bf k},t)
\end{eqnarray} 

\subsection{THE PERTURBATION APPROACH}
\subsubsection{SCHR{\"O}DINGER TO INTERACTION PICTURE}

With the above information now we are ready to study the dynamics of a many-particle system described by the density matrix $\rho$, which is specified 
by the following equation of motion
\begin{eqnarray} \label{EOM}
i \dot{\rho} & = & [H,\rho]
\end{eqnarray}
Working with the transformed operators $\rho' = U \rho U^+$ the equation of motion becomes
\begin{eqnarray}\label{transEOM}
i \dot{\rho'} & = & [H'+{ H'_d},\rho']
\end{eqnarray}
where
\begin{eqnarray} 
H'_d & = &  -\Omega{\bf  P}'.{\bf E}(t) \\ \nonumber
& = & -\sum_{nm{\bf k}} a^+_{n{\bf k}}a_{m{\bf k}}{\bf \mu}_{nm}({\bf k},t){\bf E}(t) \\
H' & = & UHU^+ \\ \nonumber
& = & \sum_{n{\bf k}} \omega_n({\bf k+K}) a^+_{n{\bf k}}a_{n{\bf k}} \label{UTH}
\end{eqnarray}
In this equation $a^+_n$ and $a_n$ are time independent and hence the commutators of $H'$ at two different times vanish. $U$ is the unitary operater
given by the Eq. \ref{uniop}. 

Until now we have been working in the Schr{\"o}dinger picture. Now we change to the interaction picture, because the equation of motion 
Eq. (\ref{EOM}) involving total Hamiltoninan simplifies to the form given by Eq. (\ref{EOMIP}), where only the perturbation term of the Hamiltonian 
$H_d$, is involved. This can be done by the use of the following relations
\begin{eqnarray}\label{wdwdt}
W(t) & = & exp \left[ i\int^t_{-\infty} H'(t') dt'\right] \nonumber \\
{{dW(t)} \over {dt}} &=& i W(t) H'(t)
\end{eqnarray}
\begin{eqnarray}
\tilde{\rho} & = &  W\rho' W^+ \nonumber \\
\tilde{a}_{m{\bf k}}  & =  &  Wa_{m{\bf k}}W^+ \nonumber \\
\tilde{a}_{m{\bf k}}(t)  & = &  a_{m{\bf k}}e^{-i\nu_m({\bf k},t)} \nonumber \\
\tilde{H}_d  =  WH'_d W^+ 
 & = & -\sum_{nm{\bf k}} \tilde{a}^+_{n{\bf k}}\tilde{a}_{m{\bf k}}{\bf \mu}_{nm}({\bf k},t){\bf E}(t).
\end{eqnarray} 
Here $\nu_m({\bf k},t) = \int^t_{-\infty} \omega_m({\bf k+K}(t'))dt'$. As shown in Appendix C Eq. (\ref{EOM}) now becomes
\begin{eqnarray} \label{EOMIP}
i \dot{\tilde{\rho}} & = & [\tilde{H}_d,\tilde{\rho}]
\end{eqnarray}
Integrating this equation we get
\begin{eqnarray}\label{IntRho}
\tilde{\rho} & = & \rho_0 + {1 \over {i}} \int^t_{-\infty} [\tilde {H}_d(t'), \rho_0] dt' \nonumber \\
&+&{1 \over {(i)^2}} \int^t_{-\infty} \int^{t'}_{-\infty} [\tilde {H}_d(t'),[\tilde {H}'_d(t''), \rho_0]] dt' dt'' \ldots
\end{eqnarray}

\subsubsection{EXPECTATION VALUES OF OPERATORS}

Our main interest lies in the single particle operators like
\begin{eqnarray}
\Theta & = & \sum_{nm{\bf k}} a^+_{n{\bf k}}a_{m{\bf k}} \Theta_{nm}({\bf k}) = \sum_{nm{\bf k}} b^+_{n{\bf k}}b_{m{\bf k}} 
\bar{\Theta}_{nm}({\bf k}) \nonumber \\
\Theta' & = &  \sum_{nm{\bf k}} a^+_{n{\bf k}}a_{m{\bf k}} \bar{\Theta}_{nm}({\bf k}) \nonumber \\
\tilde{\Theta}  & = &  \sum_{nm{\bf k}} \tilde{a}^+_{n{\bf k}}\tilde{a}_{m{\bf k}}\Theta_{nm}({\bf k+K})
\end{eqnarray}
The expectation value of any single particle operator $\Theta$ can be calculated using the relation
\begin{eqnarray} \label{trace}
<\Theta> = Tr(\rho\Theta)=Tr(\rho'\Theta') & = & Tr(\rho_0\hat{\Theta})  
\end{eqnarray}
where $\hat{\Theta}$ is 
\begin{eqnarray} \label{IntTheta}
\hat{\Theta}  &= & \tilde{\Theta}(t) + {1 \over {i}} \int^t_{-\infty} [\tilde{\Theta}(t),\tilde{H}_d(t')] dt' \nonumber \\
 &+&{1 \over {(i)^2}} \int^t_{-\infty} \int^{t'}_{-\infty} [[\tilde{\Theta}(t), \tilde{H}_d(t')],\tilde{H}_d(t'')] dt' dt''+ \ldots
\end{eqnarray}
Using the following properties of the trace 
\begin{eqnarray}
Tr(\rho_0a^+_{n{\bf k}}a_{m{\bf k}}) & = & \delta_{nm} f_n({\bf k}) \nonumber \\
Tr(\rho_0[a^+_{n{\bf k}}a_{m{\bf k}} a^+_{p{\bf k}'}a_{q{\bf k}'}]) & = & \delta_{{\bf k,k'}}\delta_{nq} \delta_{mp} f_{nm}({\bf k})
\end{eqnarray}
and combining Eqs. (\ref{IntRho}), (\ref{trace}) and (\ref{IntTheta}) the first term becomes
\begin{eqnarray}
<\Theta>_{(0)} & = & \sum_{nk} \Theta_{nn}({\bf k+K}) f_n({\bf k})
\end{eqnarray}
and the second term is 
\begin{eqnarray}\label{theta1}
<\Theta>_{(1)} &= &i\sum_{nm{\bf k}} f_{nm}({\bf k}) \Theta_{nm}({\bf k},t) e^{i\nu_{mn}({\bf k},t)} \nonumber \\
 &\times& \int^t_{-\infty}  e^{i\nu_{mn}({\bf k},t')}{\bf  \mu}_{mn}({\bf k},t').{\bf E}(t')dt'
\end{eqnarray}
The higher order terms can be written in a similar manner. Taking ${\bf E}(t) = \sum_{\beta} {\bf E}(\omega_{\beta})e^{-i\omega_{\beta}t}$, using the following 
identity
\begin{eqnarray}\label{math}
e^{S(t)}L(t)  & = &  {d \over dt}\left[ e^{S(t)} {{L(t)} \over {\dot S(t)}} \right] -e^{S(t)}{d \over dt}\left[ {{L(t)} \over {\dot S(t)}} \right]
\end{eqnarray}
and substituting 
\begin{eqnarray}
S(t) & = & i(\nu_{mn}-\omega_{\beta}t) \nonumber \\
L(t) & = & {\bf \mu}^b_{mn} \nonumber \\
\dot{S}(t) & = &i(\omega_{mn}-\omega_{\beta})
\end{eqnarray}
in Eq. (\ref{theta1}) we get
\begin{eqnarray}\label{theta1F}
<\Theta>_{(1)} &=& i\sum_{nm{\bf k}} f_{nm}({\bf k}) \Theta_{nm}({\bf k},t) e^{i\nu_{nm}({\bf k},t)}\nonumber \\
 & \times & \left\{ e^{i(\nu_{mn}({\bf k},t)-\omega_{\beta}t)} {{{\bf \mu}^b_{mn}({\bf k},t)} \over {i(\omega_{mn}({\bf \kappa})- \omega_{\beta})} } 
\right. \nonumber \\
&+&i \int^t_{-\infty} e^{i(\nu_{mn}({\bf k},t')-\omega_{\beta}t')} \\
&\times& \left. {{\delta} \over {\delta{\bf \kappa}'_c}} \left[ {{{\bf \mu}^b_{mn}({\bf k},t')} \over {(\omega_{mn}({\bf \kappa}')-\omega_{\beta})} } 
\right] {\bf E}^c(t')dt' \right\}{\bf E}^b(\omega_{\beta})  \nonumber 
\end{eqnarray}
Now expanding $E(t')$ in frequency components $E(\omega)$ and using Eq. (\ref{math}) for integrating the right side of Eq. (\ref{theta1F}) one can write 
$<\Theta>_{(1)} =  <\Theta>_{(1,0)} + <\Theta>_{(1,1)} + \ldots \ldots$ where
\begin{eqnarray} \label{Th10}
<\Theta>_{(1,0)}  & = & \sum_{nm{\bf k}} f_{nm}({\bf k}) \Theta_{nm}({\bf k},t) {{{\bf \mu}^b_{mn}({\bf k},t)} \over {(\omega_{mn}-\omega_{\beta})}} 
\nonumber \\
&\times& e^{-i\omega_{\beta}t}{\bf E}^b(\omega_{\beta}) 
\end{eqnarray}
\begin{eqnarray} \label{Th11}
<\Theta>_{(1,1)} =  i \sum_{nm{\bf k}} f_{nm}({\bf k}) \Theta_{nm}({\bf k},t) {1 \over {\omega_{mn}-\omega_{\beta}-\omega_{\gamma}}}  & \times & \nonumber \\
{\delta \over {\delta{\bf  \kappa}^c}}\left[ {{{\bf \mu}^b_{mn}({\bf k},t)} \over {\omega_{mn}-\omega_{\beta}}} \right] e^{-i(\omega_{\beta}+\omega_{\gamma})t} 
{\bf E}^b(\omega_{\beta}) {\bf E}^c(\omega_{\gamma})
\end{eqnarray}
and
\begin{eqnarray} \label{Th20}
<\Theta>_{(2,0)} &=&  \sum_{nm{\bf k}} {1 \over {\omega_{mn}-\omega_{\beta}-\omega_{\gamma}}} \nonumber \\
&\times& \left\{ {{f_{nl} \Theta_{nm}({\bf k},t) {\bf \mu}^b_{ml}{\bf  \mu}^c_{ln}} \over {\omega_{ln}-\omega_{\gamma}}} \right.\nonumber \\
&+&\left.{{f_{ml} \theta_{nm}{\bf \mu}^c_{ml}{\bf  \mu}^b_{ln}} \over {\omega_{ml}-\omega_{\gamma}}} \right\}e^{-i(\omega_{\beta}+\omega_{\gamma})t} \nonumber \\
&\times& {\bf  E}^b(\omega_{\beta}) {\bf E}^c(\omega_{\gamma})
\end{eqnarray}

\subsection{LINEAR AND SECOND ORDER SUSCEPTIBILITY}

Up to this point the things have been most general. Now in order to find the linear and non linear susceptibility $\Theta$ is replaced by the 
polarization operator ${\bf P}$. Eq. \ref{PEXP} can be written as
\begin{equation}  
<{\bf P}>  =  <{\bf P}>_I + <{\bf P}>_{II} + \ldots
\end{equation}
where
\begin{eqnarray}  
<{\bf P}^a>_I  & = &  {\chi}^{(1)}_{ab}(-\omega_{\beta},\omega_{\beta})e^{-i\omega_{\beta}t}{\bf  E}^b(\omega_{\beta}) \nonumber \\
<{\bf P}^a>_{II}  & = &  {\chi}^{(II)}_{abc}(-\omega_{\beta},-\omega_{\gamma},\omega_{\beta}, \omega_{\gamma}) \nonumber \\
&\times&e^{-i(\omega_{\beta}+\omega_{\gamma})t} {\bf E}^b(\omega_{\beta}){\bf  E}^c(\omega_{\gamma})
\end{eqnarray}
Similarly expanding the intraband current density in the powers of the electric field we get
\begin{equation}  
<{\bf J_A}> = <{\bf J_A}>_I + <{\bf J_A}>_{II} + \ldots
\end{equation}
with
\begin{eqnarray}  
<{\bf J_A}>_I  & = &   {\sigma}^{(1)}_{ab}(-\omega_{\beta},\omega_{\beta})e^{-i\omega_{\beta}t}{\bf  E}^b(\omega_{\beta})\nonumber \\
<{\bf J_A}>_{II}  & = &  {\sigma}^{(II)}_{abc}(-\omega_{\beta},-\omega_{\gamma},\omega_{\beta}, \omega_{\gamma}) \nonumber \\
&\times& e^{-i(\omega_{\beta}+\omega_{\gamma})t} {\bf E}^b(\omega_{\beta}){\bf  E}^c(\omega_{\gamma}). 
\end{eqnarray}
For clean semiconductors with filled bands the conductivity $\sigma^{(1)}$ is zero and only ${\bf P}_I$ contributes to the linear term, whereas 
to the second-order term both, ${\bf P}_{II}$ and ${\bf J}_{II}$, contribute. Now substituting $\Theta ={\bf  P}^a$ in Eq. (\ref{Th10}) we get the 
expression for the linear susceptibility
\begin{eqnarray}\label{LO}
{\chi}^{(1)}_{ab}(-\omega,\omega)  & = &  {1 \over \Omega} \sum_{nm{\bf k}} f_{nm}({\bf k})
{{{\bf r}^a_{nm}({\bf k}) {\bf r}^b_{mn}({\bf k})} \over {\omega_{mn}({\bf k})-\omega}} \nonumber \\ 
& = & {{\epsilon_{ab}(\omega) -\delta_{ab}} \over 4\pi}
\end{eqnarray}
Here $\epsilon^{ab}(\omega)$ is the $ab$ component of the dielectric tensor, and ${\bf r}_{nm}$ are the position matrix elements given by
\begin{eqnarray}
{\bf r}_{nm} = {\bf \mu}_{nm}  =  {{\bf V}_{nm}({\bf k},t) \over {i \omega_{nm}({\bf k+K})}} \, \, & {\rm if} & \,  \omega_n \ne \omega_m \nonumber \\
{\bf r}_{nm}  =  0 \, \, \,\, \, \,\, \, \,\, \, \, \, \, \,\, \, \,\, \, \,\, \, \, \, \,\, \, \,\, \, \,\, \,\, \, \,\, \, \,\, \, \,\, \, \, \, \,\, \, \,
\, \, \, & {\rm if} & \,  \omega_n = \omega_m.
\end{eqnarray}
Similarly, substituting $\Theta = {\bf P}^a$ in Eqs. (\ref{Th11}) and (\ref{Th20}) and using the following identity
\begin{eqnarray} 
{\chi}^{(2)}_{abc}(-\omega_{\beta},-\omega_{\gamma},\omega_{\beta}, \omega_{\gamma})  & = &  
{\chi}^{(II)}_{abc}(-\omega_{\beta},-\omega_{\gamma},\omega_{\beta}, \omega_{\gamma}) \nonumber \\
&+&{{i{\sigma}^{(II)}_{abc}(-\omega_{\beta},-\omega_{\gamma},\omega_{\beta}, \omega_{\gamma})} \over {\omega_{\beta} +\omega_{\gamma} }}
\end{eqnarray}
we get after some rearrangement of the terms\cite{SL,sipe93} the following contributions to the SHG susceptibility: 
the interband transitions $\chi_{\rm inter}(2\omega,\omega,\omega)$,
the intraband transitions $\chi_{\rm intra}(2\omega,\omega,\omega)$ and
the modulation of interband terms by intraband terms $\chi_{\rm mod}(2\omega,\omega,\omega)$:

\begin{eqnarray}
\chi_{\rm inter}^{abc}(2\omega,\omega,\omega) & = & {1 \over {\Omega}} \sum'_{nml\bf k} W_{\bf k} \nonumber \\
&\times&\left\{ 
{{2{\bf r}^a_{nm}\{ {\bf r}^b_{ml} {\bf r}^c_{ln}\}} \over{(\omega_{ln}-\omega_{ml})(\omega_{mn}-2\omega)}}\right. \\
&-&{1 \over (\omega_{mn}-\omega)}\left[ 
{{{\bf r}^c_{lm}\{{\bf r}^a_{mn}{\bf r}^b_{nl}\}} \over{(\omega_{nl}-\omega_{mn})}} \right.\nonumber \\ 
&-& \left. \left. {{{\bf r}^b_{nl}\{{\bf r}^c_{lm}{\bf r}^a_{mn}\}} \over{(\omega_{lm}-\omega_{mn})}}\right] \right\} \nonumber
\end{eqnarray}

\begin{eqnarray}
&&\chi_{\rm intra}^{abc}(2\omega,\omega,\omega)={1 \over {\Omega}} \sum_{{\bf k}} W_{\bf k}
\left\{ \sum'_{nml} {\omega^{-2}_{mn} \over {(\omega_{mn}-\omega)}} \right. \nonumber \\
&\times&\left[\omega_{ln}{\bf r}^b_{nl}\{{\bf r}^c_{lm}{\bf r}^a_{mn}\}-\omega_{ml}{\bf r}^c_{lm}\{ {\bf r}^a_{mn}{\bf r}^b_{nl}\}\right] \nonumber  \\
&-&8i\sum'_{nm} {1 \over{\omega^2_{mn}(\omega_{mn}-2\omega)}}{\bf r}^a_{nm}\{ {\bf r}^b_{ml}{\bf r}^c_{ln}\} \nonumber \\
&+&\left.2\sum'_{nml}{{{\bf r}^a_{nm}\{ {\bf r}^b_{ml}{\bf r}^c_{ln}\}(\omega_{ml}-\omega_{ln})} \over {\omega^2_{mn}(\omega_{mn}-2\omega)}}
\right\}
\end{eqnarray}

\begin{eqnarray}
\chi_{\rm mod}^{abc}(2\omega,\omega,\omega)&=&{1 \over {2\Omega}} \sum_{{\bf k}} W_{\bf k}
\left\{ \sum_{nml} {1 \over {\omega^2_{mn}(\omega_{mn}-\omega)}} \right. \nonumber \\
&\times&\left[\omega_{nl}{\bf r}^a_{lm}\{{\bf r}^b_{mn}{\bf r}^c_{nl}\}-\omega_{lm}{\bf r}^a_{nl}\{{\bf r}^b_{lm}{\bf r}^c_{mn}\}
\right] \nonumber \\ 
&-&\left.i\sum_{nm} {{{\bf r}^a_{nm}\{{\bf r}^b_{mn}{\bf \Delta}^c_{mn}\}} \over {\omega^2_{mn}(\omega_{mn}-\omega)}}
\right\}
\end{eqnarray}

$W_{\bf k}$ is the weight of the  ${\bf k}$ point, $n$ denotes the valence states, $m$ the conduction states and $l$ denotes all states ($l \ne m, n$).
We have used the these expressions to calculate the total susceptibility in the following example.  
\section{EXAMPLE}
As an example we present the linear and nonlinear optical spectra of an InP/GaP (110) superlattice. This material is a mono-layer SL in (110) 
direction with GaP grown on top of an InP substrate. The $zz$ component of the linear frequency dependent dielectric function 
is  given in Fig. 2(a). $\epsilon^{zz}_2(\omega)$ has major peaks at 2.3eV (B), 4eV (D) and 5.5eV (F) and minor peaks at 1eV (A), 2.75eV (C), 
4.5eV (E) and 6.5eV (G). 
These peaks in the linear optical spectra can be identified from the band structure. The calculated band structure along certain symmetry directions is 
given in Fig. 3. It can be noted from the band structure plot that InP/GaP 
is a direct band gap ($E_G$) material. The calculated band gap using the local density approximation (LDA) is $E_G = 0.6eV$\cite{scissors}. 
As can be seen from the Eq. (\ref{LO}), in order to identify of these peaks we need to look at the optical matrix elements ${\bf r}_{nm}$ for various 
pairs of band $n$ and $m$. We mark the transitions, giving the major structure in $\epsilon^{zz}_2(\omega)$, in the band structure plot. These transitions 
are labeled according to the  peak labels in Fig. 2(a).

\begin{figure}[ht]
\centerline{
\includegraphics[scale=0.5,width=\columnwidth,angle=270]{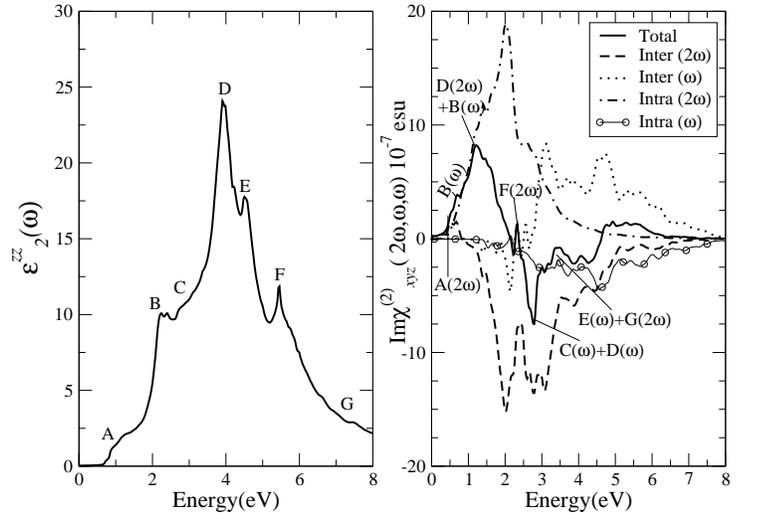}
}
\caption{(a) Imaginary part of the $zz$ component of the linear dielectric tensor. (b) Second-order susceptibility  
Im[$\chi^{(2)}_{xyz}(2\omega,\omega,\omega)$] (solid line) and different contributions to it: the ``$2 \omega$ interband term'' (dashed line), 
the ``$\omega$ interband term'' (dotted line), the ``$2 \omega$ intraband term'' (dash dotted line) and  the ``$\omega$ intraband term" (circle thin line)}
\end{figure}
\begin{figure}[ht]
\centerline{
\includegraphics[scale=0.5,width=\columnwidth,angle=270]{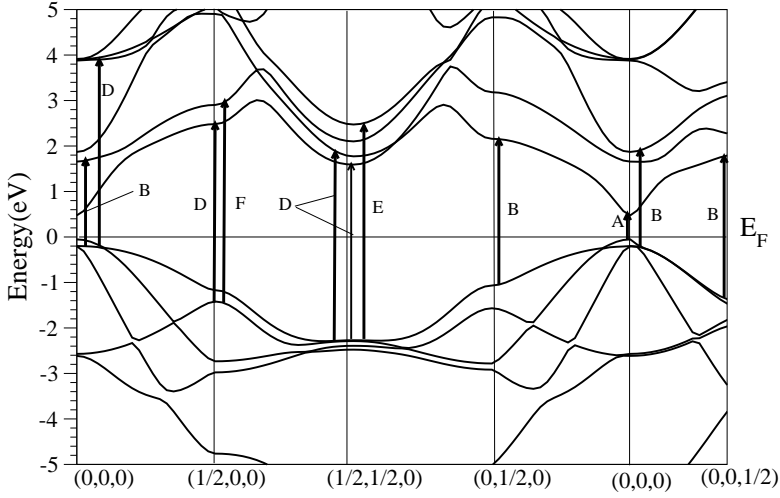}
}
\caption{ Band structure for mono-layer InP/GaP SL. }
\end{figure}
\begin{figure}[ht]
\centerline{
\includegraphics[scale=0.5,width=\columnwidth,angle=270]{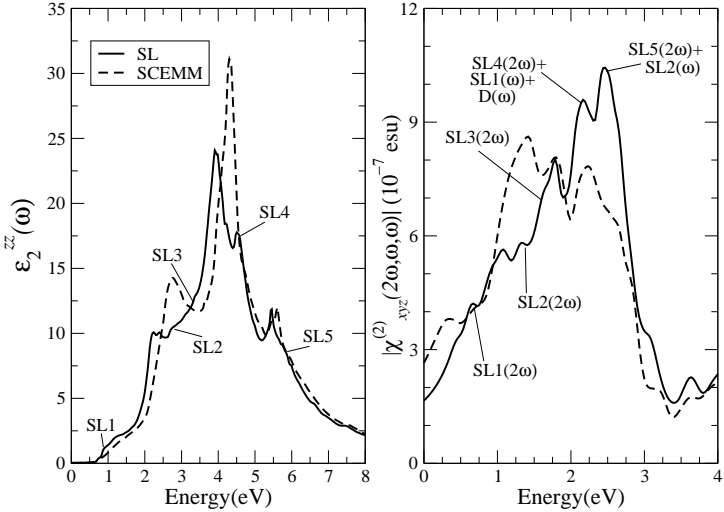}
}
\caption{ Full SL calculation (solid line) along with the SCEMM results (dashed line) for (a) the imaginary part of the linear dielectric tensor and
(b) the magnitude of the SHG susceptibility}
\end{figure}

We now go on to study the NLO properties. Different contributions to the imaginary part of $\chi^{(2)}_{xyz}(2\omega,\omega,\omega)$ are presented in 
Fig. 2(b). As can be seen the total SHG susceptibility is zero below half the band gap. 
The $2\omega$ terms start contributing at energies $\sim {1 \over 2} E_G$ and the $\omega$ terms for energy values above $E_G$. In the low energy regime 
($\le 3eV$) the SHG optical spectra is dominated by the $2\omega$ contributions. Beyond 3eV the major contribution comes from the $\omega$ terms. Unlike 
the linear optical spectra, the features in the SHG susceptibility are very difficult to identify from the band structure because of the complicated 
resonance of the $2\omega$ and $\omega$ terms. But one can make use of the linear optical spectra to identify the different resonances leading to 
various features in the NLO spectra. This analysis is performed in the present work.  The identified peaks are marked in Fig. 2,  where the
nomenclature adopted is $M(x\omega)+N(y\omega)$, which indicates that the peak comes from an  $x\omega$ resonance of the peak $M$ with the $y\omega$ 
resonance of peak {\it N} in the linear optical spectra.  For example, the hump just below 1 eV, labeled $A(2\omega)$ in the imaginary part of 
$\chi^{(2)}_{xyz}(2\omega,\omega,\omega)$ comes from the 2$\omega$ resonance of the peak labeled {\it A} in the linear optical spectra. 
The peak labeled $D(2\omega)+B(\omega)$ is coming from the 2$\omega$ resonance of the peak $D$ with $\omega$ resonance of the peak labeled $B$ in the 
$\epsilon_2(\omega)$ plot. 

Compared to the linear optical, the NLO is a much more surface/interface sensitive technique. This fact can be demonstrated by identifying the features 
coming from the interface formation. These features are referred to as SL features. These features can be pin pointed by comparing the spectra
for the SL with features appearing from the average of the two bulk materials. In order to provide a simple model for predicting the averaged bulk features 
in the optical properties on basis of its constituent materials, the effective-medium-model (EMM) and the  strain-corrected-effective-medium-model 
(SCEMM)\cite{SL} have been proposed. Comparison of the SCEMM results 
with the SL calculations are presented in Fig. 4. The SL features coming from effects like symmetry lowering are not accounted for by the SCEMM and 
are marked as SL$X$ in the figure, with $X$ representing the feature label. The small SL effects in the linear optical spectra are greatly enhanced in 
the second-order optical response. This clearly indicates the selective interface sensitivity of the NLO.  

\noindent \textbf{Acknowledgements}

We would like to thank the Austrian Science Fund for the financial support (projects P13430 and P16227). The code for calculating non-linear 
optical properties was written under the EXCITING network funded by the EU (Contract HPRN-CT-2002-00317). SS would like to thank Dr. J. K. Dewhurst
for valuable comments and suggestions. 
\section{APPENDIX A}\label{apena}
To show Eq. \ref{mu} we need to show:
\begin{equation}
{d \over dt} \left[( H(t) - \omega_n({\bf k+K}))\bar{\psi}_n({\bf k,x})\right] = 0
\end{equation}
which can be written as 
\begin{eqnarray}\label{dsedt}
[ H(t) &-& \omega_n ({\bf k+K})] {{d \bar{\psi}_n({\bf k,x})} \over dt} =  \nonumber \\
&-&{d \over dt} [H(t) - \omega_n({\bf k+K})] \bar{\psi}_n({\bf k,x})  \\
& = & e{\bf E}(t).\left[{{\delta \omega_n({\bf k+K})} \over {\delta {\bf k}}}+(i\nabla -{\bf K})\right] \bar{\psi}_n({\bf k,x}) \nonumber
\end{eqnarray}
substituting Eq. \ref{215} in Eq. \ref{dsedt} can be written as 
\begin{eqnarray}\label{219}
-&i&\sum'_{m}\omega_{mn}({\bf k+K})\mu_{mn}.{\bf E(t)}(\bar{\psi}_m({\bf k,x})e^{i{\bf K.x}}) \\
& = & e{\bf E}(t).\left[{{\delta \omega_n({\bf k+K})} \over {\delta {\bf k}}}+i\nabla \right] (\bar{\psi}_n({\bf k,x})e^{i{\bf K.x}})\nonumber
\end{eqnarray}
The prime indicates that $m \ne n$. Note that here we have used the fact 
$i\nabla (\bar{\psi}_ne^{i{\bf K.x}}) =e^{i{\bf K.x}} (i\nabla - {\bf K}) \bar{\psi}_n $. The projection of LHS of this Eq. on 
$\bar{\psi}_l({\bf k,x}) e^{i{\bf K.x}}$ is   
\begin{eqnarray}
-\int \bar{\psi}^*_l({\bf k,x}) e^{-i{\bf K.x}}&i&\sum'_{m}\omega_{mn}({\bf k+K})\mu_{mn}\nonumber \\
&.&{\bf E(t)}\bar{\psi}_m({\bf k,x})e^{i{\bf K.x}}d{\bf x}
\end{eqnarray} 
For the case with $l = n$ ($m \ne n$), this is zero. The RHS of Eq. \ref{219} is also zero because
\begin{eqnarray}
i\int \bar{\psi}^*_n({\bf k,x})e^{-i{\bf K.x}} \nabla \bar{\psi}_n({\bf k,x})e^{i{\bf K.x}}d{\bf x} &=& -{\bf V}_{nn}({\bf k},t) \\
&=&-{ {\delta \omega_n({\bf k+K})}  \over {\delta {\bf k}} } \nonumber 
\end{eqnarray}
For the case $l \ne n$ the LHS of Eq. \ref{219} is
\begin{eqnarray}
&-&i\int \bar{\psi}^*_l({\bf k,x})e^{-i{\bf K.x}} \sum_m \omega_{mn} \mu_{mn}.{\bf E}(t) (\bar{\psi}_m({\bf k,x})e^{i{\bf K.x}}) \nonumber \\
= &-& i\omega_{ln}({\bf k+K})\mu_{ln}.{\bf E}(t) 
\end{eqnarray}
The RHS of Eq. \ref{219} in this case is $-e{\bf V_{ln}.E}(t)$. Thus the condition for two sides to be equal is
\begin{equation}
{\bf \mu}_{mn}({\bf k+K},t) =  {{{\bf V}_{mn}({\bf k},t)} \over {i\omega_{mn}({\bf k+K})}}
\end{equation}

\section{APPENDIX B}\label{apenb}
In this section we confirm Eq.\ref{ja}. We start with the relation
\begin{eqnarray}
<{\bf P}> &=& Tr(\rho{\bf P}) = Tr(\rho'{\bf P'}) \nonumber \\ 
&=& {1 \over \Omega} Tr \left[\rho' \sum_{nm{\bf k}}a^+_{n{\bf k}}a_{m{\bf k}}\mu_{nm} \right]
\end{eqnarray}
Now taking the time derivative of the expectation value of ${\bf P}$ we get
\begin{eqnarray}
{ {d<P^a>} \over dt } & = & {1 \over \Omega} Tr \left( {d \over dt} \left[\rho' \sum_{nm{\bf k}}a^+_{n{\bf k}}a_{m{\bf k}}\mu_{nm} \right] \right) \nonumber \\
& = & {1 \over \Omega} Tr 
\left( 
\left[ 
{ {d \rho'} \over dt } \sum_{nm{\bf k}} a^+_{n{\bf k}} a_{m{\bf k}} \mu_{nm} 
\right]
\right. \nonumber \\ 
&+& \left.\left[ \rho' \sum_{nm{\bf k}}a^+_{n{\bf k}}a_{m{\bf k}} { {d \mu_{nm}} \over dt} \right] 
\right).
\end{eqnarray}
Using the equation of motion \ref{EOMIP} this can be written as:
\begin{eqnarray} \label{expP}
{ {d<P^a>} \over dt } & = & {1 \over {i\Omega}}
Tr \left( [H',\rho'] \sum_{nm{\bf k}}a^+_{n{\bf k}}a_{m{\bf k}}\mu^a_{nm} \right) \nonumber \\
& + & {1 \over {i\Omega}} Tr \left( [H'_d,\rho'] \sum_{nm{\bf k}}a^+_{n{\bf k}}a_{m{\bf k}}\mu^a_{nm} \right) \nonumber \\
& + & {1 \over {\Omega}} Tr \left( \rho'\sum_{nm{\bf k}}a^+_{n{\bf k}}a_{m{\bf k}} 
{ {\delta \mu^a_{nm}}  \over {\delta {\bf K^b}} } \right) {d{\bf K^b} \over dt} 
\end{eqnarray}
further using the property of trace and operator relations
\begin{equation}
Tr([A,B]C) = Tr(B[C,A]),
\end{equation}
\begin{equation}
H' = \sum_{n {\bf k}}\omega_{n}({\bf k+K})a^+_{n{\bf k}}a_{m{\bf k}},
\end{equation}
\begin{equation}
[a^+_{n{\bf k}}a_{m{\bf k}},H'] = - \omega_{nm}({\bf k+K})a^+_{n{\bf k}}a_{m{\bf k}},
\end{equation}
and
\begin{eqnarray}
[a^+_{n{\bf k}}a_{m{\bf k}},a^+_{p{\bf k'}}a_{q{\bf k'}}] &=& a^+_{n{\bf k}}a_{q{\bf k}}\delta_{mp}\delta_{{\bf kk'}}\nonumber \\
&-&a^+_{p{\bf k}}a_{m{\bf k}}\delta_{nq} \delta_{{\bf kk'}}
\end{eqnarray}
the first term in Eq. \ref{expP} becomes
\begin{eqnarray} \label{ja}
& = & {1 \over {i\Omega}} Tr\left( \rho' \left[ \sum_{nm{\bf k}}a^+_{n{\bf k}}a_{m{\bf k}}\mu^a_{nm}, 
\sum_{n{\bf k}}a^+_{n{\bf k}}a_{n{\bf k}} \omega_n \right] \right) \nonumber \\
& = & {1 \over \Omega} \left\{ 
-Tr\left(\rho'\sum_{n{\bf k}} {\bf V}^a_{nn}({\bf k+K}) a^+_{n{\bf k}} a_{n{\bf k}}\right) \right. \nonumber \\
&+& \left.Tr\left( \rho' \sum_{nm{\bf k}}a^+_{n{\bf k}}a_{m{\bf k}} (\omega_n\delta_{nm} - \mu^a_{nm}.E(t)) \right)  \right\}
\end{eqnarray}
where the last term in this equation is ${\bf J_A}$


\section{APPENDIX C}\label{apenc}
We need to prove
\begin{equation}
i \dot{\tilde{\rho}} = [\tilde{H}_d,\tilde{\rho}]
\end{equation}
The LHS can be written as 
\begin{eqnarray}
i \dot{\tilde{\rho}} & = & i {d \over dt}{\tilde{\rho}}  = i {d \over dt}[{W\rho'W^+}] \nonumber \\
& = & i {dW \over dt} \rho' W^+ + i W {{d\rho'} \over dt} W^+ + i W\rho' {{dW^+} \over dt} 
\end{eqnarray}
Now using Eqs. \ref{transEOM} and \ref{wdwdt} this becomes
\begin{eqnarray}
&-&WH'\rho'W^+ +  W[H'+H'_d,\rho']W^+ + W\rho'H'W^+ \nonumber \\
&=&W[H'+H'_d,\rho']W^+ - W[H',\rho']W^+ \nonumber \\
&=& W[{H}_d,\rho']W^+ = [\tilde{H}_d,\tilde{\rho}].
\end{eqnarray}


\begin{thebibliography}{10}
\bibitem{scissors}
The LDA is known to underestimate the bandgaps. The usual procedure is to use
  the scissors operator to correct the bandgap for calculating the optical
  response of the material. In order to keep things simple, in the present work
  no scissors correction is used. The scissors corrected results can be seen in
  the Ref. 10.
\end{thebibliography}

\end{document}